\documentclass[preprint,twocolumn]{aastex631}
\usepackage[utf8]{inputenc}
\usepackage{savesym}
\savesymbol{tablenum}
\usepackage{siunitx}
\restoresymbol{SIX}{tablenum}
\usepackage{amsmath}
\usepackage{natbib}
\usepackage{float}
\usepackage{appendix}
\usepackage{multirow}
\usepackage{booktabs}
\usepackage{hyperref}
\usepackage{comment}
\newcommand{\CellWithForceBreak}[2][c]{\begin{tabular}[#1]{@{}c@{}}#2\end{tabular}}

\begin{document}

\title{A High-Resolution Spectroscopic Survey of Directly Imaged Companion Hosts: \\
III. Characterization of the Cold Imaged Planet Hosts AF Lep A and $\epsilon$ Indi A}

\author[0000-0003-3708-241X]{Aneesh Baburaj} \affiliation{Center for Interdisciplinary Exploration and Research in Astrophysics, Northwestern University, 1800 Sherman Ave, Evanston, IL 60201, USA}

\author[0000-0002-9936-6285]{Quinn M. Konopacky} \affiliation{Department of Astronomy \& Astrophysics, University of California, San Diego, La Jolla, CA 92093, USA} 

\author[0000-0002-9807-5435]{Christopher A. Theissen} \affiliation{Department of Astronomy \& Astrophysics, University of California, San Diego, La Jolla, CA 92093, USA} 

\author[0000-0002-6618-1137]{Jerry W. Xuan}
\altaffiliation{51 Pegasi b Fellow}
\affiliation{Division of Geological \& Planetary Sciences, California Institute of Technology, Pasadena, CA 91125, USA}

\author[0000-0003-0398-639X]{Roman Gerasimov} \affiliation{Department of Physics and Astronomy, University of Notre Dame, Nieuwland Science Hall, Notre Dame, IN 46556, USA}

\author[0000-0002-9803-8255]{Kielan K. W. Hoch} \affiliation{Space Telescope Science Institute, 3700 San Martin Drive, Baltimore, MD 21218, USA}
 

\correspondingauthor{Aneesh Baburaj}
\email{ababuraj@northwestern.edu}

\begin{abstract}
JWST has enabled the measurement of carbon, oxygen, and sulfur abundances in the atmospheres of directly imaged planets. Interpretation of these abundances from a planet formation standpoint requires the corresponding abundances for the host star. In this work, we present detailed characterizations of the cold imaged planet hosts AF Lep A and $\epsilon$ Indi A using high-resolution Gemini/GHOST spectra. We derive the atmospheric parameters $T_{\rm eff}$ and $\log{g}$ using two different approaches, revealing differences in $T_{\rm eff}$ up to $\sim260\,$K. The derived parameters are subsequently incorporated in measurement of 16 elemental abundances (C, O, Na, Mg, Si, S, K, Ca, Sc, Ti, Cr, Mn, Fe, Ni, Zn, Y) and several abundance ratios. Utilizing both the spectral fit and the equivalent width methods, we find solar C/O, C/S and O/S ratios ($<1.5\sigma$) for AF Lep A and $\epsilon$ Indi A. We compare our measured abundances and their ratios with those of the planets AF Lep b and $\epsilon$ Indi Ab, with the elevated abundances for the planets relative to their host stars strongly indicating formation by core-accretion pathways.  
\end{abstract}

\section{Introduction \label{sec:intro}}
Direct imaging surveys have led to the discovery of several dozen planets over the past two decades, with this planetary population uniquely characterized by their Jovian to super-Jovian masses and wide range of separations (3--5000\,au) from their primary (e.g., \citealt{nielsen2019, vigan2021, zjzhang2021, hinkley2023}). These diverse properties imply that there is likely no single pathway for the formation of imaged planets, with both bottom-up core accretion or top-down gravitational/disk instability being viable. While past models of core accretion struggled to reach the critical core mass for runaway accretion ($\sim10\,M_\oplus$) beyond $\sim5\,$au \citep{dodson2009}, recent theoretical advances in pebble accretion are able to form planets up to 10$\,M_{\rm Jup}$ even at $\sim40\,$au before dissipation of the disk (e.g., \citealt{emsenhuber1,emsenhuber2}). However, the masses and separations of imaged planets are also consistent with the lower mass end of formation predictions by gravitational instability \citep{kratter2016}.

Studies of exoplanet atmospheres are a potential avenue to break the degeneracy between the various formation scenarios, with atmospheric abundance ratios like the carbon-to-oxygen ratio (C/O) proposed as formation fingerprints \citep{oberg2011,madhu2011}. This led to a slew of C/O measurements for directly imaged planets (e.g., \citealt{2013Sci...339.1398K, 2021AJ....162..290R, 2021A&A...648A..59P, 2023AJ....166...85H, xuan2024b, balmer2025}). However, there are considerable degeneracies involved in the use of C/O ratio as a formation tracer (e.g., \citealt{2019ARA&A..57..617M, turrini2021, pacetti2022}). For instance, \cite{turrini2021} show that all scenarios of combined gas+solid or gas-only accretion during the runaway phase of gas giant formation lead to nearly stellar final (post-migration) atmospheric C/O ratio for the planet, regardless of the initial core formation location. Furthermore, \cite{pacetti2022} note that similar volatility of both C and O reduce the diagnostic ability of the C/O ratio, making the atmospheric C/O alone insufficient for constraining the accretion history of the planet without additional measurements of its atmospheric metallicity. Additionally, observations of directly imaged planet and brown dwarf atmospheres have measured C/O ratios similar to solar values (e.g., \citealt{xuan2022,xuan2024b,hsu2024,costes2024,picos2024,snellen2025}). This has prompted investigation into other elemental ratios as tracers of formation histories, with subsequent work proposing volatile-to-sulfur ratios (C/S and O/S) as tracers of a planet's accretion history \citep{crossfield2023}. Highly refractory elements like soldium (Na), potassium (K), iron (Fe), and silicon (Si) can be used to measure volatile-to-refractory ratios, with these ratios tracing both formation and accretion histories \citep{sb2021b,turrini2021,pacetti2022,chachan2023}.

More recently, elevated metallicities of exoplanet atmospheres have been proposed as a more reliable tracer of planet-like formation. Several wide separation planets show hints of super-stellar metallicities (e.g., AF Lep b: \citealt{balmer2025, denis2025}, HR 8799 planets: \citealt{wang2023, nasedkin2024, balmer2025b}), suggesting an extensive history of solid accretion \citep{jiwang2025, ruffio2026, xuan2026}. This has been further reinforced by the revised mass-metallicity relations \citep{chachan2025}, where the metallicity plateau even at super-Jupiter masses hint at accretion of several tens to hundred of Earth masses of solids as being a characteristic signature of planet-like formation. However, elevated atmospheric metallicities do not reveal the extent of solid accretion in the planetary atmosphere. These results have led to observational studies seeking to measure the enrichment of both volatile (C, O) and refractory elements (S) in exoplanet atmospheres, with the latter being tracers of solid accretion in the planetary atmosphere \citep{xuan2026}. The super-stellar abundances of C, O, and S for the HR 8799 planets \citep{ruffio2026,xuan2026} as well as super-solar elemental abundances for Jupiter \citep{Wong2004,Li2020} and Saturn \citep{Briggs1989,Fletcher2009,Atreya2018} in our Solar System indicate this might be a promising avenue.

However, determination of elemental enrichment in the context of planet formation requires the corresponding elemental abundances for the host star. Direct imaging surveys primarily observe young ($<$100 Myr) stars \citep{nielsen2019,vigan2021} which are often poorly characterized compared to their older counterparts owing to their high rotation velocities ($v\sin{i}\,>10\,$kms$^{-1}$) which leads to significant blending of spectral lines. As JWST expands our ability to measure detailed elemental abundances in exoplanet atmospheres, it is imperative that we measure the corresponding abundances for the host star. This has been the motivation behind our previous work \citep{aneesh2025,aneesh26}, which has measured abundances of up to 16 elements (C, O, Na, Mg, Si, S, K, Ca, Sc, Ti, Cr, Mn, Fe, Ni, Zn, Y) for 10 directly imaged companion host stars, including primaries of multi-planetary systems like HR 8799, YSES-1, and $\beta$ Pic. Our methods have enabled abundance measurements for rotators up to $v\sin{i}\sim120\,$kms$^{-1}$, obtaining abundance precisions $\sim0.1\,$dex, which is sufficient for constraining formation pathways (e.g., \citealt{sb2021b}). 

In this work, we utilize methods developed in \cite{aneesh26} to obtain detailed elemental abundances for directly imaged planet hosts AF Lep A and $\epsilon$ Indi A. AF Lep b is one of closest Jupiter analogs both in terms of mass ($\sim3.5\,M_{\rm Jup}$) and orbital separation ($\sim9\,$au, \citealt{franson2023,derosa2023,mesa2023}), while $\epsilon$ Indi Ab is the coldest known imaged planet ($T_{\rm eff}\sim275\,$K, \citealt{matthews2024}). Thus, these planets have been of great interest among the imaging community as their atmospheres can be characterized using direct spectroscopy. We determine the stellar atmospheric parameters of the host stars, and subsequently use these to obtain abundances of C, O, Na, Mg, Si, S, K, Ca, Sc, Ti, Cr, Mn, Fe, Ni, Zn, Y at precisions 0.1--0.2 dex. Section \ref{sec:targets} describes our targets in detail, after which we briefly detail our observing plan and data reduction (Section \ref{sec:obs}). Our methods and the changes compared to \cite{aneesh26} are explained in Section \ref{sec:methods}, with the results of our analysis highlighted in Section \ref{sec:results}. In Section \ref{sec:discussion}, we perform literature comparison for our stellar parameters and abundances, with the latter subsequently compared to the abundances for the planets. Finally, we summarize our conclusions in Section \ref{sec:conclusions}.

\section{Targets \label{sec:targets}}
\subsection{AF Lep A}
AF Lep A is a young, F8V star at a distance of 26.843 $\pm$ 0.014 pc \citep{gaiaedr3}. A confirmed member of the $\beta$ Pictoris moving group (BPMG), it has a well-constrained age of 18--26 Myr \citep{malo2014,bell2015, miret2020}. It also has a literature mass estimate of 1.20 $\pm$ 0.06 $\,M_\odot$ \citep{kervella2022}. A combination of astrometric and direct imaging techniques led to the discovery of the Jovian planet AF Lep b, with a dynamical mass of 3.75 $\pm\,0.5\,M_{\rm Jup}$, orbiting the primary at a separation of 8.98$^{+0.15}_{-0.08}$ au (initial discovery by \citealt{franson2023,derosa2023,mesa2023}; planetary parameters from \citealt{balmer2025}). At the time of its discovery, AF Lep was the lowest mass Jupiter analog orbiting its host star at Solar System scales, leading to extensive work on its atmospheric characterization. Studies of its atmosphere using photometry \citep{zj2023,palma2024,franson2024}, low-resolution \citep{balmer2025}, and high-resolution spectroscopy \citep{denis2025}, utilizing both forward modeling (with self-consistent atmospheric model grids) and retrieval analyses, have consistently found super-solar metallicities and solar C/O ratio ([M/H] = 0.75 $\pm$ 0.25, C/O = 0.55 $\pm$ 0.10; values from \citealt{balmer2025}). However, there are no measurements of carbon, oxygen, sulfur abundances for AF Lep A.

\subsection{$\epsilon$ Indi A}
$\epsilon$ Indi A is a K5V star in the solar neighborhood, located at a distance of 3.638 $\pm$ 0.001 pc \citep{gaiaedr3}. It also has a mass estimate of 0.76 $\pm$ 0.04$\,M_\odot$ obtained using radii derived from interferometry and empirical mass-radius relations \citep{demory2009}. The primary has distant brown dwarf binary companions ($\epsilon$ Indi Ba/Bb, \citealt{scholz2003, mccaugh2004}), both of which have well-constrained dynamical masses between 50--70$\,M_{\rm Jup}$ \citep{chen2022}. \cite{chen2022} also used stellar activity indicators to estimate an age of 3.5$^{+0.8}_{-1.0}$ Gyr for the system. \cite{feng2019} used radial velocity and astrometry data to propose the presence of a super-Jupiter around the primary, which was eventually confirmed using mid-infrared imaging with JWST/MIRI \citep{matthews2024}. Further imaging of $\epsilon$ Indi Ab with JWST/NIRSpec and JWST/MIRI imaging has found dynamical masses of 6.5--7.5$\,M_{\rm Jup}$ \citep{sanghi2026, matthews2026}. \cite{sanghi2026} also estimated super-solar metallicities ([M/H]$\,\sim0.7$) for the planet. However, these atmospheric studies have been exclusively using photometric data and forward modeling with self-consistent atmospheric model grids. As a solar neighborhood star, $\epsilon$ Indi A has several measurements of elemental abundances in the literature. Most recently, \citep{2022AJ....164...87K} measured [C/H] = -0.19 $\pm$ 0.05, [O/H] = -0.11 $\pm$ 0.04, and [S/H] = -0.26 $\pm$ 0.06 using archival FEROS data. It must be noted that their abundances use \cite{asplund2021} as solar reference.

\section{Observations and Data Reduction \label{sec:obs}}
The spectra were obtained using the Gemini High-resolution Optical SpecTrograph \citep[GHOST;][]{ghost1,ghost2} at the Gemini South Observatory. GHOST is an echelle spectrograph with a red arm and a blue arm separated by a dichroic at 530 nm. The detector corresponding to the blue arm is sensitive to wavelengths 347--542 nm, while the red arm detector is sensitive to wavelengths 520--1060 nm. The GHOST high-resolution mode allows us to obtain $R\sim76,000$ spectra of our targets across both arms. Our observations seek to mitigate the probability of chance events (e.g., cosmic rays) affecting our observations, hence we divide our observations into multiple integrations of equal exposure times. The exact details of our spectral observations are given in Table \ref{tab:obs}.

\begin{deluxetable*}{ccccc}
\tablecaption{Observing schedule for targets in this work\label{tab:obs}}
\tablewidth{0pt}
\tablehead{
\colhead{Target} & \CellWithForceBreak{Number of \\ exposures} & \CellWithForceBreak{Integration Time \\ per exposure (s)} & \CellWithForceBreak{Observation date \\ (UT)} & \CellWithForceBreak{Total Int. \\ Time (s)} 
}
\startdata
{AF Lep A}  & {8} & {240}  & {December 13, 2025} & {1920} \\ \hline
\multirow{2}{*}{$\epsilon$ Indi A} & {4} & {35} & {October 23, 2025}   & \multirow{2}{*}{210} \\
{} & {2} & {35} & {October 24, 2025}   & {} \\ \hline
\enddata
\end{deluxetable*}

The data reduction is performed using the DRAGONS (Data Reduction for Astronomy from Gemini Observatory North and South) data reduction pipeline \citep{2023RNAAS...7..214L,simpson_2026_19055103}, a Python package provided by Gemini Observatory for the reduction of all astronomical data. Data reduction by the GHOST instrument pipeline within DRAGONS follows the same procedure discussed in \cite{aneesh26}. Similar to the latter work, we do not apply sky correction for tellurics as the telluric features are important for precise wavelength solutions in our spectral fits. Flux calibration using a spectrophotometric standard is not required as we perform an order-specific continuum normalization of all spectra before any further analysis.

\section{Methods \label{sec:methods}}
\subsection{Determination of Atmospheric Parameters \label{subsec:atmosparam}}

Initially, we utilize a similar approach to our previous work in \cite{aneesh26} for the determination of stellar parameters like temperature, gravity, and metallicity. The telluric features in order \#55 are used to constrain the telluric parameters. The H$\alpha$ line of the Balmer series (6562.79 \r{A}) is used to estimate the effective temperature ($T_{\rm eff}$). Metallicity ($\rm [M/H]$) is constrained using the Na \textsc{i} D doublet, and wavelength regions from 5600--5750 \r{A} and 6000--6200 \r{A}. Mg \textsc{i}b triplet (5150--5200 \r{A}) provides estimates of the surface gravity ($\log{g}$). This is followed by multi-order fits to refine $T_{\rm eff}$, $\log{g}$, and $\rm [M/H]$. Application of this approach to the spectrum of AF Lep A leads to $T_{\rm eff}$ = 5814 $\pm$ 100 K, $\log{g}$ = 4.36 $\pm$ 0.18, [M/H] = $-0.16 \pm 0.09$ dex. However, spectroscopic determination of atmospheric parameters can have systematic offsets of $\sim$350 K in $T_{\rm eff}$ \citep{2021A&A...653A..90M,2026A&A...708L..17H}. Thus, we complement our spectroscopic analysis with atmospheric parameters determined from evolutionary models. Atmospheric analysis using measured photometry and comparing them to synthetic photometry predicted from evolutionary models to determine temperature and gravity have been used in existing literature (e.g., \citealt{kolecki2021,reggiani2020,reggiani2021,reggiani2022}).

\begin{deluxetable*}{c|cccc}
\tablecaption{Parameters for evolutionary model analysis\label{tab:evomodel}}
\tablewidth{0pt}
\tablehead{
\colhead{Parameter} & \colhead{AF Lep A} & \colhead{$\epsilon$ Indi A} & \colhead{References} 
}
\startdata
{Tycho-2 $B$} & {$6.954 \pm 0.007$} &  {$6.069 \pm 0.003$}  & {1} \\ 
{Tycho-2 $V$} & {$6.362 \pm 0.006$} &  {$4.833 \pm 0.002$}  & {1} \\ 
{2MASS $J$}   & {$5.268 \pm 0.027$} &  {$2.894 \pm 0.292$}  & {2} \\ 
{2MASS $H$}   & {$5.087 \pm 0.026$} &  {$2.349 \pm 0.214$}  & {2} \\ 
{2MASS $K_s$} & {$4.926 \pm 0.021$} &  {$2.237 \pm 0.240$}  & {2} \\
{Distance (pc)}  & {26.843 $\pm$ 0.014} &  {3.638 $\pm$ 0.001}  & {3} \\ 
{System age} & {18--26 Myr} & {3.5$^{+0.8}_{-1.0}$ Gyr}  & {AF Lep A: 4,5,6; $\epsilon$ Ind A: 7} \\ 
{Stellar metallicity ([M/H])\tablenotemark{a}} & {-0.18 $\pm$ 0.09} & {-0.14 $\pm$ 0.08} & {This work} \\
{Stellar mass ($M_{\odot}$)} & {$1.20 \pm 0.06$} & {$0.76 \pm 0.04$}  & {AF Lep A: 8; $\epsilon$ Ind A: 9} \\ 
\enddata
\tablenotetext{}{NOTE -- All magnitudes are in Vega mags}
\tablenotetext{a}{The \texttt{PARSEC v1.2S} evolutionary tracks use \cite{caffau2011} as solar reference, as opposed to \cite{asplund2009} used in our spectroscopic determination of stellar metallicity. Hence, we apply a correction for the different solar references before using our measured stellar metallicity for evolutionary model analyses.}
\tablerefs{(1) \cite{Tycho2}, (2) \cite{2MASS}, (3) \cite{gaiaedr3}, (4) \cite{malo2014}, (5) \cite{bell2015}, (6) \cite{miret2020}, (7) \cite{chen2022}, (8) \cite{kervella2022}, (9) \cite{demory2009}}
\end{deluxetable*}

We utilize the \texttt{PARSEC v1.2S} evolutionary tracks \citep{2012MNRAS.427..127B,2014MNRAS.444.2525C,2015MNRAS.452.1068C,2017ApJ...835...77M}, as implemented in the online tool \texttt{CMD 3.9\footnote{https://stev.oapd.inaf.it/cgi-bin/cmd}} to obtain evolutionary model estimates of the $T_{\rm eff}$ and $\log{g}$. \texttt{CMD} interpolates between \texttt{PARSEC} evolutionary tracks at different masses and metallicities and can generate isochrone tables with several stellar parameters at a range of masses for a range of user-specified stellar metallicities and ages. Stellar parameters include temperature, gravity, and absolute magnitudes in various filters. The isochrone tables generated by CMD are in the form of ascii files and are easily utilized for our analysis. We generate these isochrone tables with metallicity [M/H] $\in[-0.5,0.5]$. 

As a member of BPMG, the stellar age of AF Lep A is well-constrained (18--26 Myr; \citealt{malo2014,bell2015,miret2020}). We have apparent magnitudes in the Tycho-2 \citep{Tycho2} and 2MASS \citep{2MASS} filters, and the distance to the system from Gaia EDR3 \citep{gaiaedr3} (Table \ref{tab:evomodel}). Subsequently, we utilize a Markov Chain Monte Carlo (MCMC) approach using the \texttt{emcee} package \citep{emcee} to compare the known stellar age, metallicity from the spectroscopic analysis (corrected for the different solar reference; refer Table \ref{tab:evomodel}), and filter magnitudes (after converting from absolute to apparent magnitude) against the isochrone tables and obtain posteriors of the $T_{\rm eff}$ and $\log{g}$. Our MCMC sampling utilizes 20 chains of 3000 steps each, with the first 500 discarded as burn-in. We keep wide priors for $T_{\rm eff}\in(2500 \,\rm K,8000\, \rm K)$ and $\log{g}\in(3.4,5.1)$. The log-likelihood chains are used to compute the median and standard deviation. This approach yields $T_{\rm eff}$ = 6076$^{+180}_{-92}\,$K and $\log{g}$ = 4.31$^{+0.10}_{-0.03}$. We adopt the $T_{\rm eff}$ and $\log{g}$ from the evolutionary model analysis and spectroscopically determined metallicity ([M/H] = $-0.16 \pm 0.09$) as the stellar atmospheric parameters for AF Lep A.

We perform similar atmospheric analysis for $\epsilon$ Indi A. The initial spectroscopic analysis yields $T_{\rm eff}$ = 4560 $\pm$ 42 K, $\log{g}$ = 4.60 $\pm$ 0.11, [M/H] = $-0.12 \pm 0.08$. We supplement our spectroscopy with similar analysis using the \texttt{PARSEC v1.2S} evolutionary tracks and the parameters from Table \ref{tab:evomodel}, which yields $T_{\rm eff}$ = 4687$^{+39}_{-29}\,$K and $\log{g}$ = 4.61$^{+0.04}_{-0.02}$. Incorporating both the spectroscopic and evolutionary model estimates, we adopt $T_{\rm eff}$ = 4687$^{+39}_{-29}$ K, $\log{g}$ =  4.61$^{+0.04}_{-0.02}$, and [M/H] = $-0.12 \pm 0.08$ as the stellar atmospheric parameters for $\epsilon$ Indi A.

Our dual spectroscopic and evolutionary model approach for measurement of stellar atmospheric parameters reveals systematics of up to $\sim260\,$K in the determination of the effective temperature $T_{\rm eff}$, with the gravity being majorly consistent across both approaches. This combined analyses allows us to account for the effects of temperature systematics while determining the elemental abundances for our host stars.

\subsection{Determination of Elemental Abundances}
We utilize methods from our previous work \citep{aneesh26} to measure carbon and oxygen abundances via a spectral template fitting method. Synthetic grids are generated using PySME, a Python wrapper for the Spectroscopy Made Easy (SME) spectral analysis code \citep{valenti1996,piskunov2017}. We utilize the MARCS stellar atmosphere grids \citep{marcs2008} and atomic line-lists from the Vienna Atomic Line Database (VALD3; \citealt{piskunov2015, 2015PhyS...90e4005R}) to generate model grids (\textit{MARCS-C/O}) with varying carbon and oxygen abundances using stellar parameters determined in Section \ref{subsec:atmosparam}. While the $T_{\rm eff}$ and $\log{g}$ determined using evolutionary models are adopted as the fiducial stellar parameters, we include estimates from both the spectroscopic and evolutionary model analyses while determining $T_{\rm eff}$ and $\log{g}$ ranges for these model grids. Each model grid has fixed metallicity, [C/H]$\,\in\,$[$-$0.4,0.4], and [O/H]$\,\in\,$[$-$0.5,0.5] in steps of 0.1$\,dex$. The solar abundances are taken from \cite{asplund2009}. The carbon and oxygen lines used for these measurements are the same as those in \cite{aneesh26}. We do not use all lines for abundance measurement for each target, depending on the strength of the specific C \textsc{i}/O \textsc{i} feature and the general fit of our model to the spectral order with the feature. NLTE effects for the O \textsc{i} lines are corrected using 3D NLTE to 1D LTE corrections from \cite{amarsi2019}.

Recent work on exoplanet atmospheres has identified sulfur as an element of interest in estimating the amount of solid accretion into the planetary atmosphere \citep{crossfield2023,fu2024,xuan2026}, leading to the measurement of this element for several directly imaged companions \citep{ruffio2026,xuan2026,aneesh26b}. Hence, we also seek to expand our spectral template method to measure sulfur abundances. For this purpose, we create custom grids (henceforth called \textit{MARCS-C/S}) using the same prescription as described above, but with varying carbon and sulfur abundances. Model grids for all targets have $T_{\rm eff}$ and $\log{g}$ from our previous spectroscopic/evolutionary model analyses, fixed metallicity, [C/H]$\,\in\,$[$-$0.4,0.4], and [S/H]$\,\in\,$[$-$0.5,0.5], with the latter abundances varying in steps of 0.1$\,dex$. These grids also use solar abundances derived from \cite{asplund2009}.

The sulfur (S \textsc{i}) features in order \# 51 (6743--6744 \r{A} and 6757 \r{A}) are used for determining the sulfur abundance. The sulfur features chosen have negligible contributions from carbon atomic lines. Regardless, we fix the carbon abundance to the value determined using the \textit{MARCS-C/O} grid fits while estimating the sulfur abundance. As an additional check to the robustness of our determined abundances, we use the \textit{MARCS-C/S} grid to fit for carbon abundances (at fixed [S/H]) for the same orders for which abundances were previously measured using the \textit{MARCS-C/O} grid. We recover the same values as determined using the \textit{MARCS-C/O} grid.

In addition to spectral fitting, we use the equivalent width (EW) approach to measure the abundances of carbon, oxygen, sulfur, and 13 additional elements (Na, Mg, Si, K, Ca, Sc, Ti, Cr, Mn, Fe, Ni, Zn, Y). The approach is again similar to \cite{aneesh26}; the spectral analysis software MOOG \citep{1973ApJ...184..839S} and the Kurucz-ATLAS9 model atmospheres generated using the BasicATLAS framework \citep{2023AJ....165....2L} are used to convert the EW measurements to corresponding abundances for each spectral species. We use two different drivers within MOOG for the measurement of our metal abundances. For lines without significant blending, we use the `abfind' driver that force-fits abundances to single line EW to obtain abundance values corresponding to those spectral lines. For lines with significant rotational blending, we use \texttt{numpy.trapz} to determine the EW of the blended spectral feature. The NIST Atomic Line Database \citep{Kramida2024NISTASD} provides us with the individual spectral lines present in each blended feature. The `blends' driver in MOOG is then used to obtain abundance corresponding to the spectral species of interest. `blends' uses the Kurucz-ATLAS9 model provided to calculate the EW for the lines not corresponding to the species of interest in the blended spectral feature. The residual EW encapsulates the contribution from the desired species and is converted to an equivalent abundance. For each target, uncertainties in the EW abundance account for the line-to-line abundance scatter, as well as errors in $T_{\rm eff}$ and $\log{g}$. For species with only a single line, we use the measurement errors in the EW as a substitute for the line-to-line abundance scatter. As we are estimating the absolute abundance of a species, the metallicity ([M/H]) uncertainties have only a minor effect on the result, with the majority of the metal abundance uncertainty coming from $T_{\rm eff}$ and $\log{g}$. For O \textsc{i}, we use 3D NLTE to 1D LTE corrections from \cite{amarsi2019}. For K \textsc{i}, we use 1D NLTE to 1D LTE corrections from \cite{reggiani2019}. In their paper, the latter show that both 1D NLTE and 3D NLTE corrections for K \textsc{i} have comparable values. Thus, the use of 1D NLTE (rather than 3D NLTE) for potassium does not significantly affect our derived abundances.

\section{Results \label{sec:results}}

\begin{deluxetable*}{lccccccc}
\tabletypesize{\small} 
\tablecaption{Equivalent width elemental abundance corresponding to various spectral species\label{tab:ewabundance}}
\tablewidth{0pt}
\tablehead{
\colhead{Species} & \colhead{AF Lep A} & \colhead{} & \colhead{} & \colhead{$\epsilon$ Indi A} & \colhead{} & \colhead{} & \colhead{Solar values} \\
\colhead{} & 
\colhead{log(N)} & \colhead{$\sigma$} & \colhead{\# lines}  &
\colhead{log(N)} & \colhead{$\sigma$} & \colhead{\# lines}  &
\colhead{}
}
\startdata
C \textsc{i}   & 8.39 & 0.13 & 2 & 8.21 & 0.07 & 1 & 8.43 \\
O \textsc{i}   & 8.79 & 0.14 & 1\tablenotemark{\footnotesize{a}} & 8.55 & 0.08 & 4 & 8.69 \\
Na \textsc{i}  & 5.90 & 0.19 & 1 & 6.19 & 0.05 & 1 & 6.24 \\
Mg \textsc{i}  & 7.42 & 0.09 & 3 & 7.42 & 0.12 & 3 & 7.60 \\
Si \textsc{i}  & 7.39 & 0.22 & 4 & 7.39 & 0.08 & 11 & 7.51 \\
S \textsc{i}   & 7.05 & 0.10 & 1 & 6.97 & 0.18 & 1 & 7.12 \\
K \textsc{i}\tablenotemark{\footnotesize{b}} & 4.85 & 0.15 & 1 & 4.83 & 0.06 & 2 & 5.03 \\
Ca \textsc{i}  & 6.13 & 0.15 & 6 & 6.20 & 0.13 & 12 & 6.34 \\
Sc \textsc{ii} & 3.00 & 0.05 & 2 & 3.01 & 0.06 & 6 & 3.15 \\
Ti \textsc{i}  & 4.86 & 0.15 & 3 & 4.74 & 0.08 & 8 & 4.95 \\
Ti \textsc{ii} & 4.75 & 0.11 & 7 & 4.78 & 0.05 & 8 & 4.95 \\
Cr \textsc{i}  & 5.47 & 0.15 & 3 & 5.44 & 0.11 & 8 & 5.64 \\
Cr \textsc{ii} & 5.40 & 0.09 & 4 & 5.45 & 0.07 & 2 & 5.64 \\
Mn \textsc{i}  & 5.32 & 0.25 & 3 & 5.38 & 0.05 & 4 & 5.43 \\
Fe \textsc{i}  & 7.34 & 0.17 & 45 & 7.38 & 0.10 & 76 & 7.50 \\
Fe \textsc{ii} & 7.33 & 0.12 & 12 & 7.32 & 0.11 & 11 & 7.50 \\
Ni \textsc{i}  & 6.09 & 0.17 & 4 & 6.06 & 0.07 & 9 & 6.22 \\
Zn \textsc{i}  & 4.37 & 0.14 & 2 & 4.38 & 0.04 & 3 & 4.56 \\
Y \textsc{ii}  & 2.03 & 0.07 & 3 & 2.06 & 0.07 & 2 & 2.21 \\
\enddata
\tablenotetext{}{All abundances are given as absolute [X/H] + 12}
\tablenotetext{a}{O I triplets blended into a single feature are counted as singular lines}
\tablenotetext{b}{K \textsc{i} values reported after applying NLTE corrections from \cite{reggiani2019}}
\end{deluxetable*}

\subsection{AF Lep A} 
As discussed in Section \ref{subsec:atmosparam}, the stellar atmospheric parameter analysis for AF Lep A yields $T_{\rm eff}$ = 6076$^{+180}_{-92}$ K, $\log{g}$ =  4.31$^{+0.10}_{-0.03}$, and [M/H] = -0.16 $\pm$ 0.09. These values are used to generate \textit{MARCS-C/O} and \textit{MARCS-C/S} grids with $T_{\rm eff}\,\in[5700, 6300]$ K, $\log{g}\,\in[4.0,5.0]$, and fixed metallicity ([M/H] = -0.16). The former set of grids have [C/H]$\,\in[-0.4,0.4]$ and [O/H]$\,\in[-0.5,0.5]$, while the latter have [C/H]$\,\in[-0.4,0.4]$ and [S/H]$\,\in[-0.5,0.5]$. 

\begin{figure*}
    \centering
    \plottwo{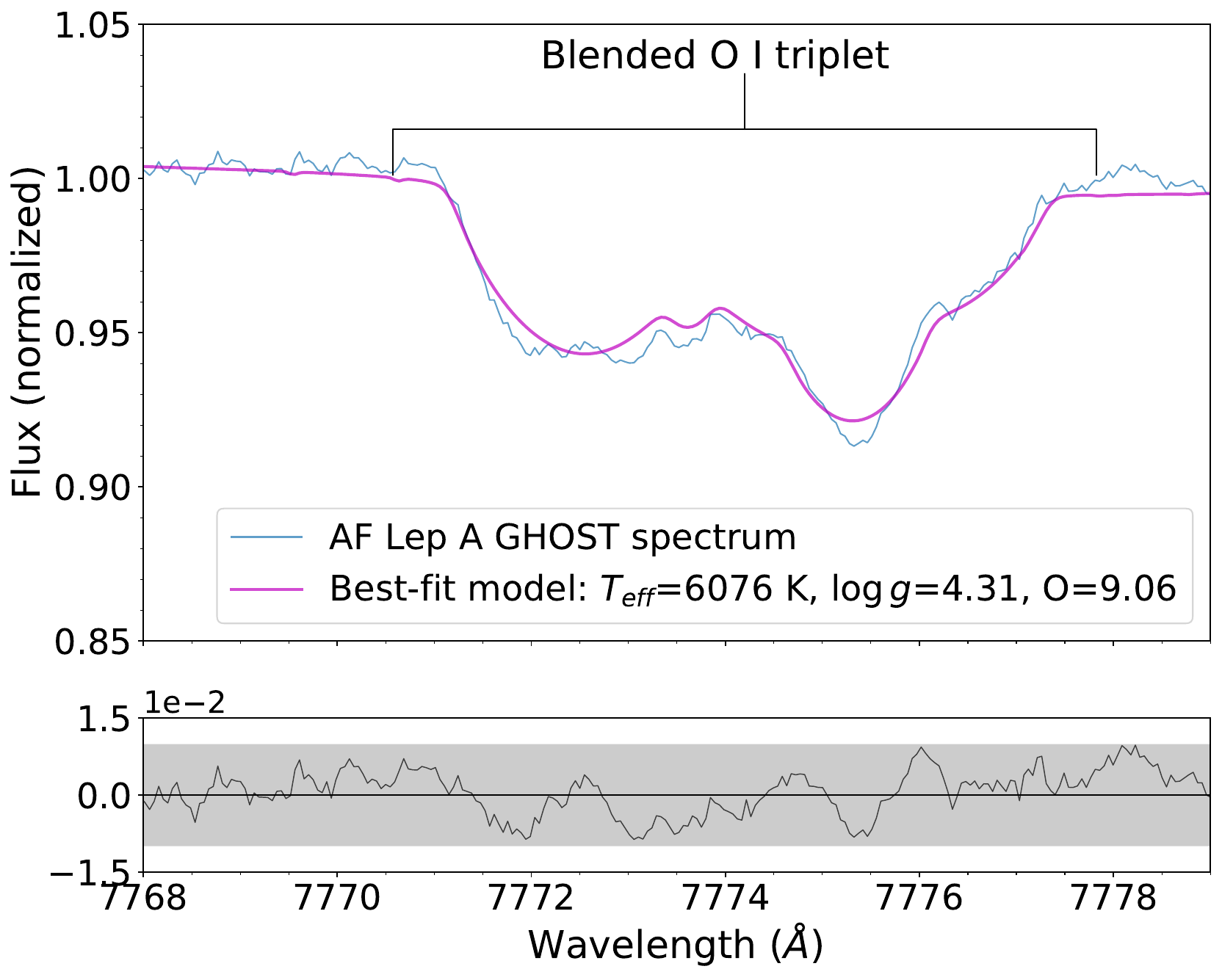}{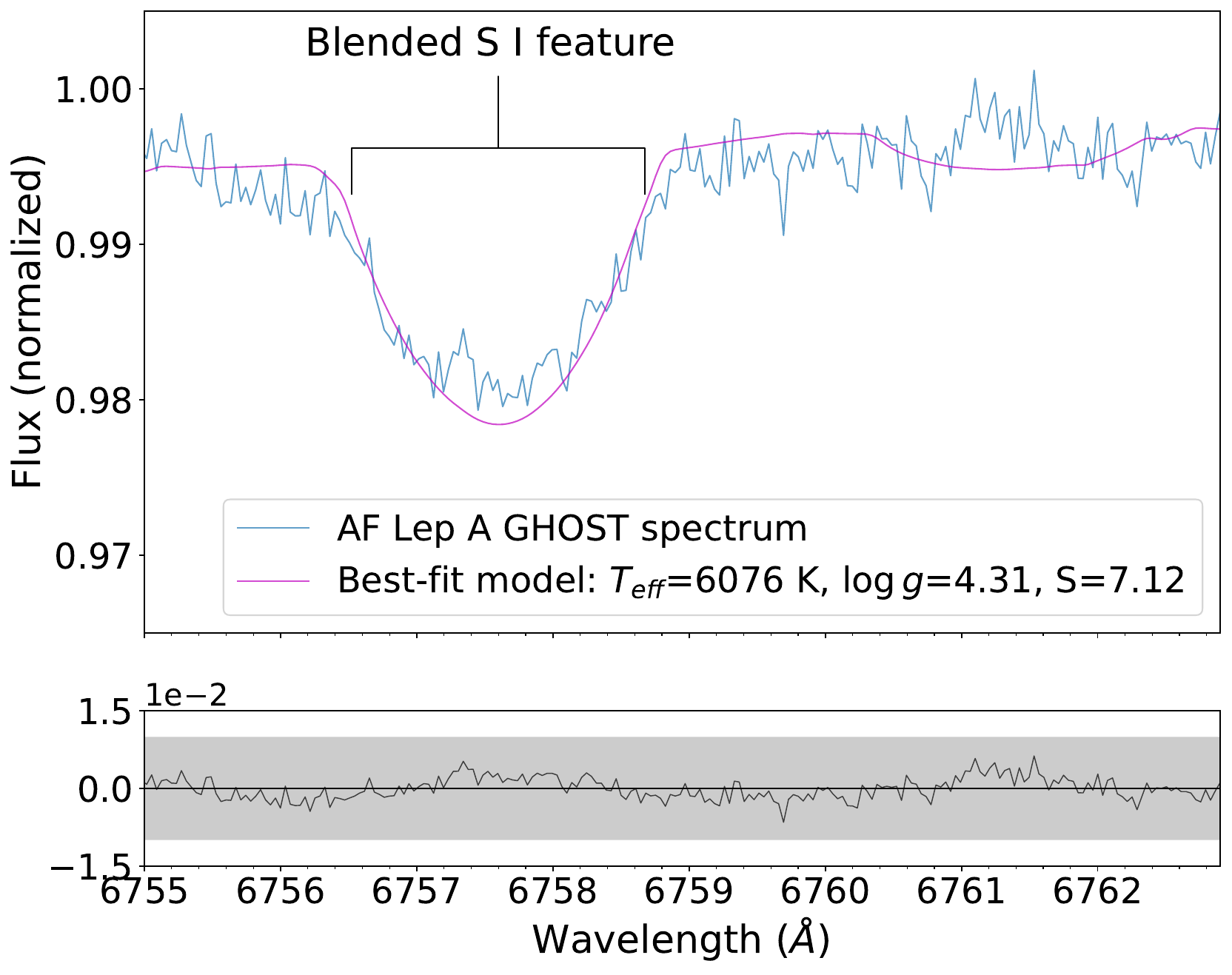}
    \caption{Best-fit model (magenta) to the GHOST spectrum of AF Lep A (cyan), shown for (\textit{Left}) the blended O \textsc{i} line at 7771--75 \r{A} and (\textit{Right}) the blended S \textsc{i} feature at 6757 \r{A}. The residuals between the data and the model are plotted in black and other noise limits are shown in gray. Fitting the \textit{MARCS-C/O} grid to the O \textsc{i} triplet feature gives a best-fit $\log{\epsilon_O}$ = 9.06. After applying NLTE corrections, we get $\log{\epsilon_O}$ = 8.84. The \textit{MARCS-C/S} grid fit to the S \textsc{i} spectral feature gives a best-fit $\log{\epsilon_S}$ = 7.12}
    \label{fig:aflepspec}
\end{figure*}

\begin{figure*}
    \centering
    \includegraphics[width=0.8\linewidth]{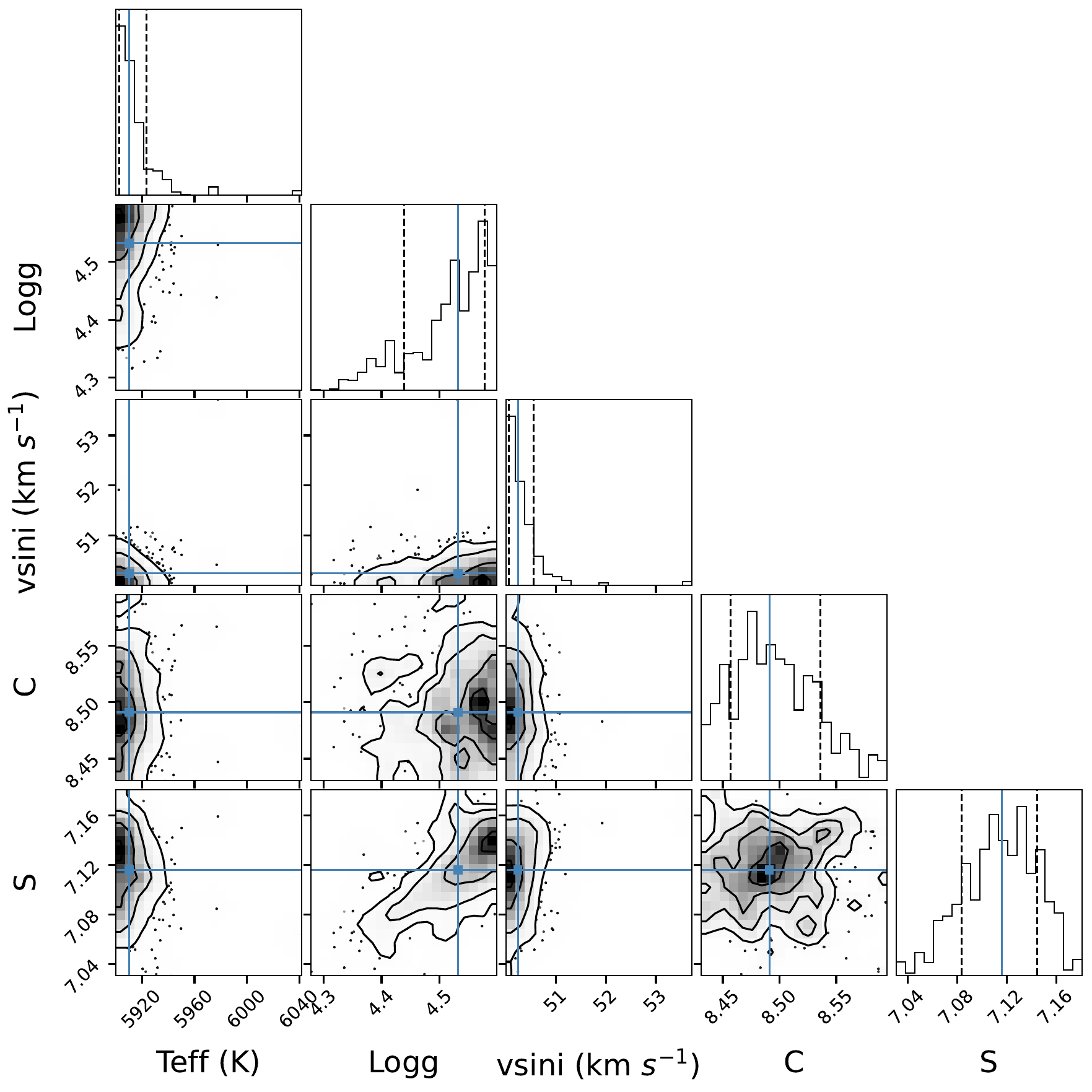}
    \caption{Posteriors for \textit{MARCS}--\textit{C/S} grid fit to spectral order \#51 with the S \textsc{i} feature at 6757 \r{A}. The marginalized posteriors are shown along the diagonal. The blue lines represent the 50 percentile, and the dotted lines represent the 16 and 84 percentiles. The subsequent covariances between all the parameters are in the corresponding 2-D histograms. We obtain a best-fit $\log{\epsilon_S}$ = 7.12 $\pm$ 0.03. Note that the latter uncertainty values solely represent the fitting uncertainties. Other parameters (including carbon abundance) are not well-constrained fitting this order.}
    \label{fig:corneraflepcs}
\end{figure*}

For the carbon abundance, we perform MCMC fits with 1500 steps and 110 walkers over three carbon orders, with the first 1000 discarded as burn-in, giving us [C/H] = 0.05 $\pm$ 0.05. Oxygen is measured using single-order MCMC fits involving order \#44 with the 7771-75 \r{A} O \textsc{i} triplet, yielding [O/H] = 0.37$^{+0.14}_{-0.11}$. After NLTE corrections of -0.22 dex from \cite{amarsi2019}, we obtain a final [O/H] = 0.15$^{+0.14}_{-0.11}$. The sulfur measurement is derived using single order fits to order \#51, giving [S/H] = 0.00 $\pm$ 0.09. These abundances give us a spectral fit C/O = 0.44$^{+0.15}_{-0.12}$, C/S = $22.91 \pm 5.43$, and O/S = $52.48^{+20.11}_{-17.17}$. The spectral fits of our model grids to the 7771--75 \r{A} O \textsc{i} triplet and the S \textsc{i} feature at 6757 \r{A} are shown in Figure \ref{fig:aflepspec}. Figure \ref{fig:corneraflepcs} shows the posteriors for the spectral fit of the \textit{MARCS}-\textit{C/S} grid to the latter S \textsc{i} feature. All abundance ratios are detailed in Table \ref{tab:abundratio}.

Meanwhile, the equivalent width approach gives [C/H] = -0.04 $\pm$ 0.13, [O/H] = 0.10 $\pm$ 0.14 (after NLTE corrections), and [S/H] = -0.07 $\pm$ 0.10. All three agree with their spectral fit values within $<1.5\sigma$ significance. Computing the abundance ratios, we get C/O = 0.40 $\pm$ 0.18, C/S = 21.88 $\pm$ 8.26, O/S = 54.95 $\pm$ 21.77, all solar within 1$\sigma$. Among other species, Sc \textsc{ii}, Cr \textsc{ii}, and Y \textsc{ii} are sub-solar at $>2\sigma$. However, given that our Cr \textsc{i} is solar within 1$\sigma$, the under-abundance for Cr \textsc{ii} could be due to underestimated measurement errors. 

\subsection{$\epsilon$ Indi A}
The stellar parameters adopted for $\epsilon$ Indi A are $T_{\rm eff}$ = 4687$^{+39}_{-29}$ K, $\log{g}$ =  4.61$^{+0.04}_{-0.02}$, and [M/H] = -0.12 $\pm$ 0.08, which are used to generate custom abundance grids with $T_{\rm eff}\,\in[4500, 5100]$ K, $\log{g}\,\in[4.0,5.0]$, and [M/H] = -0.12. The oxygen is determined using single-order fits to the 7771-75 \r{A} O \textsc{i} triplet, giving [O/H] = -0.07$^{+0.06}_{-0.05}$ after NLTE correction of -0.04 dex. For the carbon measurement, the oxygen is kept fixed, with the carbon abundance allowed to vary freely. Subsequent spectral fits over two carbon orders yield [C/H] = -0.15 $\pm$ 0.06. A sulfur abundance of [S/H] = -0.11 $\pm$ 0.20 is obtained from single order spectral fits. Hence, we derive abundance ratios of C/O = 0.46$^{+0.09}_{-0.08}$, C/S = 18.62 $\pm$ 8.95, and O/S = 40.74$^{+19.59}_{-19.34}$ from the spectral fitting approach. We obtain a sub-solar [C/H] ($>2\sigma$), and solar C/O, C/S, and O/S ratios. All three abundance ratios are shown in Table \ref{tab:abundratio}.

Equivalent width measurements give us [C/H] = -0.22 $\pm$ 0.07, [O/H] = -0.14 $\pm$ 0.08, and [S/H] = -0.15 $\pm$ 0.18, leading to abundance ratios of C/O = 0.46 $\pm$ 0.11, C/S = 17.38 $\pm$ 7.67, and O/S = 38.02 $\pm$ 17.12. The [C/H] is sub-solar ($>3\sigma$), while C/O, C/S, and O/S are solar within 1$\sigma$. All three abundance ratios also agree with those derived from the spectral fitting approach. Of the other species, K \textsc{i}, Sc \textsc{ii}, Ti \textsc{i}, Ti \textsc{ii}, Cr \textsc{ii}, Ni \textsc{i}, Zn \textsc{i}, and Y \textsc{ii} deviate from solar by $>2\sigma$. However, our quoted K \textsc{i} abundance errors do not account for any probable systematics in NLTE corrections.

\begin{deluxetable}{c|c|c|c}
\tablecaption{Abundance ratios for AF Lep A and $\epsilon$ Indi A \label{tab:abundratio}}
\tablewidth{0pt}
\tablehead{
\colhead{Abundance ratio} & \colhead{AF Lep A} & \colhead{$\epsilon$ Indi A} & \colhead{Solar}}
\startdata
C/O (spectral fit) & $0.44^{+0.15}_{-0.12}$ & $0.46^{+0.09}_{-0.08}$ & \multirow{2}{*}{0.55}  \\
C/O (eq. width)    & $0.40 \pm 0.18$        & $0.46 \pm 0.11$        &       \\
\hline
C/S (spectral fit) & $22.91 \pm 5.43$ & $18.62 \pm 8.95$ & \multirow{2}{*}{20.42} \\
C/S (eq.width)     & $21.88 \pm 8.26$ & $17.38 \pm 7.67$ &       \\
\hline
O/S (spectral fit) & $52.48^{+20.11}_{-17.17}$ & $40.74^{+19.59}_{-19.34}$ & \multirow{2}{*}{37.15} \\
O/S (eq.width)     & $54.95 \pm 21.77$ & $38.02 \pm 17.12$ &  
\enddata
\end{deluxetable}

\begin{deluxetable*}{c|c|cccccc}
\tabletypesize{\scriptsize} 
\tablecaption{Literature comparisons \label{tab:litcompare}}
\tablewidth{0pt}
\tablehead{\colhead{Target}    & \colhead{Work}     & \colhead{$T_\mathrm{eff}$ (K)}    & \colhead{$\log{g}$ (cgs)}     & \colhead{[M/H]}     & \colhead{[C/H]} & \colhead{[O/H]} & \colhead{[S/H]}
}
\startdata
\multirow{6}{*}{AF Lep A}  & \multirow{2}{*}{This work} & \multirow{2}{*}{6076$^{+180}_{-92}$} & \multirow{2}{*}{4.31$^{+0.10}_{-0.03}$} & \multirow{2}{*}{-0.16 $\pm$ 0.09} & 0.05 $\pm$ 0.05 \tablenotemark{a} &  0.15$^{+0.14}_{-0.11}$ \tablenotemark{a} & 0.00 $\pm$ 0.09 \tablenotemark{a}\\ 
  & & & & & $-0.04 \pm 0.13$ \tablenotemark{b} & $0.10 \pm 0.14$ \tablenotemark{b} & $-0.07 \pm 0.10$ \tablenotemark{b}\\ 
  & \cite{zj2023} & 5997 $\pm$ 147 & 4.30 $\pm$ 0.05 & -0.27 $\pm$ 0.31 \tablenotemark{c} & & & \\
  & \cite{soubiran2022} & 6227 $\pm$ 102 & 4.59 $\pm$ 0.22 & 0.28 $\pm$ 0.16 \tablenotemark{c} & & & \\  
  & \cite{casa2011} & 6123 $\pm$ 80 & 4.33 & -0.05 & & & \\ \hline
\multirow{7}{*}{$\epsilon$ Indi A}  & \multirow{2}{*}{This work} & \multirow{2}{*}{4687$^{+39}_{-29}$} & \multirow{2}{*}{4.61$^{+0.04}_{-0.02}$} & \multirow{2}{*}{-0.12 $\pm$ 0.08} & -0.15 $\pm$ 0.06 \tablenotemark{a} &  -0.07$^{+0.06}_{-0.05}$ \tablenotemark{a} & -0.11 $\pm$ 0.20 \tablenotemark{a}\\ 
  & & & & & -0.22 $\pm$ 0.07 \tablenotemark{b} & -0.14 $\pm$ 0.08 \tablenotemark{b} & -0.15 $\pm$ 0.18 \tablenotemark{b}\\ 
  & \cite{soubiran2024} & 4676 $\pm$ 35 & 4.59 $\pm$ 0.01 & -0.13 $\pm$ 0.03 \tablenotemark{c} & & & \\ 
  & \cite{2022AJ....164...87K} \tablenotemark{d} & 4682 $\pm$ 40 & 4.60 $\pm$ 0.02 & -0.12 $\pm$ 0.07 \tablenotemark{c} & $-0.16 \pm 0.05$ & $-0.11 \pm 0.04$ & $-0.26 \pm 0.06$ \\
  & \cite{casa2011} & 4731 $\pm$ 80 & 4.67 & 0.01 & & & \\
  & \cite{sousa2008} & 4754 $\pm$ 89 & $4.45 \pm 0.19$ & $-0.20 \pm 0.04$ & & & \\ \hline
\enddata
\tablenotetext{a}{Abundances obtained using spectral fitting}
\tablenotetext{b}{Abundances obtained using equivalent width}
\tablenotetext{c}{Value corresponds to [Fe/H] and not [M/H]}
\tablenotemark{d}{The elemental abundances from \cite{2022AJ....164...87K} have been corrected as they use \cite{asplund2021} as solar reference, as opposed to \cite{asplund2009} used in this work.}

\end{deluxetable*}

\section{Discussion \label{sec:discussion}}
\subsection{Comparison to literature measurements}
\subsubsection{AF Lep A}
AF Lep A has multiple measurements of stellar parameters in the literature, with measurements by \cite{casa2011}, \cite{soubiran2022}, and \cite{zj2023}. From Table \ref{tab:litcompare}, we see that our measurements of $T_{\rm eff}$ and $\log{g}$ agree with literature within a maximum of 2$\sigma$. Our measured sub-solar metallicity is well within the range of previous measurements across the literature.

While AF Lep A has previous abundance measurements for Fe, Mg, and Ca \citep{zj2023}, no measurements are available for C, O, or S. However, prior studies involving chemical tagging have determined that co-natal stars (i.e., stars that form in the same cluster) have the same chemical composition (e.g., \citealt{ness2018, andrews2019}). This suggests that we can compare the abundances of AF Lep A to other stars from the beta Pic moving group. In particular, we use abundance measurements for the slow rotating F-type star HD 181327 from \cite{reggiani2024}. As a member of BPMG, its abundances have been suggested as a reliable tracer for protoplanetary disk composition in the $\beta$ Pictoris planetary system. We compare our abundances in Table \ref{tab:aflep_metals}. Abundance comparisons between AF Lep A and HD 181327 indicate that both spectral fit and equivalent width carbon agree within 1.5$\sigma$. Sulfur abundance from both approaches agrees with the corresponding value for HD 181327. The greatest tension is in the oxygen abundance, where both equivalent width and spectral fit measurements for AF Lep A differ by $>1\sigma$ relative to HD 181327. This could possibly be due to differences in NLTE corrections, which can lead to systematics in oxygen abundance of the order $\sim0.18$ dex \citep{reggiani2024}.   

\begin{deluxetable*}{c|cc|cc}
\tabletypesize{\small} 
\tablecaption{AF Lep A elemental abundance comparisons \label{tab:aflep_metals}}
\tablewidth{0pt}
\tablehead{
\colhead{Comparison target} & \colhead{Element} & \colhead{Measured value from prev.} & \colhead{This work} & {(AF Lep A)} \\
\colhead{} & \colhead{} & \colhead{literature} & \colhead{Spectral fit} & \colhead{Eq. width}
}
\startdata
\multirow{2}{*}{AF Lep A}    & [Fe/H]  & $-0.27 \pm 0.31$ & - & $-0.16 \pm 0.16$ \\
                             & [Mg/H]  & $-0.11 \pm 0.21$ & - & $-0.18 \pm 0.09$ \\
(\citealt{zj2023})              & [Ca/H]  & $-0.32 \pm 0.26$ & - & $-0.21 \pm 0.15$ \\ 
\hline
\multirow{2}{*}{HD 181327}\tablenotemark{a,b}  & [C/H] & $-0.05 \pm 0.06$ &  $0.05 \pm 0.05$ & $-0.04 \pm 0.13$\\
                                             & [O/H] & $-0.10 \pm 0.06$ & $0.15^{+0.14}_{-0.11}$ & $0.10 \pm 0.14$\\
(\citealt{reggiani2024})                     & [S/H] & $ 0.08 \pm 0.26$ & $0.00 \pm 0.09$ & -$0.07 \pm 0.10$\\
\enddata
\tablenotetext{a}{HD 181327 and AF Lep A are both members of the beta Pictoris moving group}
\tablenotetext{b}{The abundances for HD 181327 from \cite{reggiani2024} have been corrected to account for their different solar reference \citep{asplund2021} as opposed to the \cite{asplund2009} used in this work.}
\end{deluxetable*}

\subsubsection{$\epsilon$ Indi A}
As a slow-rotating ($v\sin{i}<2\,$kms$^{-1}$) main-sequence star just 3.6 pc from the Sun, $\epsilon$ Indi A has been studied extensively, with a small subset of results highlighted in Table \ref{tab:litcompare}. Summarizing the numerous studies on the atmospheric properties of this star (e.g., \citealt{santos2001,santos2004,ramirez2005,sousa2008,casa2011,adi2012,boyajian2012,ramirez2013,tsantaki2013,delgado2015,delgado2017,luck2018,soto2018,hojjat2019,hj2022,soubiran2022,2022AJ....164...87K,soubiran2024,perdenwitz2024}), we find $T_{\rm eff}$ = 4550--4900 K, $\log{g}$ = 4.25--4.68, [M/H] = -0.23--0.01. Our derived parameters fall well within this range.

$\epsilon$ Indi A also has previous abundance measurements, with the most recent by \cite{2022AJ....164...87K}. The latter used archival FEROS spectra to obtain abundances of several elements critical to planet formation, including C, O, Na, Mg, Si, S, K, and Fe. We compare our abundances to their values in Table \ref{tab:epsind_metals}, finding that our abundances agree with their values within a maximum of 1.5$\sigma$ for both the spectral fit and equivalent width approaches. 

In addition, it is also apparent that our abundance uncertainties are higher than those measured by the former, especially for sulfur. This is primarily due to the weak sulfur feature, with even a small discrepancy in equivalent width leading to large abundance errors. 

\begin{deluxetable*}{c|c|cc}
\tabletypesize{\small} 
\tablecaption{Abundance comparison between this work and \citealt{2022AJ....164...87K} for $\epsilon$ Indi A \label{tab:epsind_metals}}
\tablewidth{0pt}
\tablehead{
\colhead{Element} & \colhead{\cite{2022AJ....164...87K}} & \colhead{This work} & \colhead{} \\
\colhead{} & \colhead{} & \colhead{Spectral fit} & \colhead{Eq. width}}
\startdata
$\rm [C/H]$   & $-0.16 \pm 0.05$ & $-0.15^{+0.06}_{-0.06}$ & $-0.22 \pm 0.07$\\
$\rm [O/H]$   & $-0.11 \pm 0.04$ & $-0.07^{+0.06}_{-0.05}$ & $-0.14 \pm 0.08$\\
$\rm [Na/H]$  & $-0.22 \pm 0.09$ & - & $-0.05 \pm 0.05$\\
$\rm [Mg/H]$  & $-0.08 \pm 0.07$ & - & $-0.18 \pm 0.12$\\
$\rm [Si/H]$  & $ 0.06 \pm 0.06$ & - & $-0.12 \pm 0.08$\\
$\rm [S/H]$   & $-0.26 \pm 0.06$ & $-0.11 \pm 0.20$ & $-0.15 \pm 0.18$\\
$\rm [K/H]$   & $-0.12 \pm 0.05$ & - & $-0.20 \pm 0.06$\\
$\rm [Fe/H]$  & $-0.16 \pm 0.07$ & - & $-0.13 \pm 0.10$\\
\enddata
\tablenotetext{}{NOTE -- The abundances from \cite{2022AJ....164...87K} have been corrected to account for their different solar reference \citep{asplund2021} instead of the \cite{asplund2009} used in this work.}
\end{deluxetable*}

\subsection{Stellar vs Planetary Abundances}
As mentioned in Section \ref{sec:intro}, the C/O ratio has been widely used as a planet formation diagnostic. Theoretical models indicate that formation via gravitational instability would lead to stellar C/O ratios owing to the non-separation of gas and solids. Meanwhile, the pebble/core accretion formalism is proposed to lead to a range of C/O ratios depending on planet formation location relative to the various snowlines, such as H$_2$O, CO$_2$, CH$_4$, and CO (e.g., \citealt{oberg2011,piso2015,sb2021b,chachan2023}). In Table \ref{tab:companiontable}, we present both the planetary abundances as well as the stellar abundances measured in this work.

While AF Lep b has similar C/O ratio to AF Lep A, the large uncertainties prevent us from relying solely on this diagnostic. A recent JWST study on AF Lep b \citep{xuan2026b} measured super-stellar atmospheric abundances for carbon, oxygen, and sulfur, strongly indicating that this planet formed by pebble/core-accretion. Furthermore, their sub-stellar C/S and O/S ratios point toward formation between the H$_2$S and CO snowlines. A detailed discussion on the formation of AF Lep b will be included in \cite{xuan2026b} and is beyond the scope of this paper. For $\epsilon$ Indi Ab, both \cite{matthews2026} and \cite{sanghi2026} estimate C/O = 2.5$\times\,$solar ($\sim$1.38), which is significantly higher compared to $\epsilon$ Indi A. The atmospheric C/O, in conjunction with their estimated metallicity ([M/H]$\sim$0.7 dex) provides strong evidence for formation by pebble/core-accretion. However, we caution that the abundances for $\epsilon$ Indi Ab are from self-consistent model fits to photometric data. Future work using JWST/NIRSpec and JWST/MIRI spectroscopy (GO 8714, PI: J. Xuan) will provide more accurate estimates of the planet's abundances and hence its formation history.

\begin{deluxetable*}{ccccccc}
\tablecaption{Planetary and stellar abundances\label{tab:companiontable}}
\tablewidth{0pt}
\tablehead{\colhead{Planet} & \colhead{Mass ($M_\mathrm{Jup}$)} & \CellWithForceBreak{Planet \\ $\rm [M/H]$} & \CellWithForceBreak{Planet \\ C/O} & \CellWithForceBreak{Stellar \\ $\rm [M/H]$} & \CellWithForceBreak{Stellar \\ C/O \tablenotemark{\footnotesize{1}}} & \colhead{References}} 
\startdata
AF Lep b \tablenotemark{\footnotesize{2}} & 3.75$\,\pm\,0.5$ & $0.67^{+0.06}_{-0.07}$ & $0.43 \pm 0.02$ & $-0.16 \pm 0.09$ & $0.44^{+0.15}_{-0.12}$ & {1, 2, 5} \\
$\epsilon$ Indi Ab & $7.63^{+0.73}_{-0.70}$ & $\sim0.7$ & $\sim1.38$ & $-0.12 \pm 0.08$ & 0.46$^{+0.09}_{-0.08}$ & {3, 4, 5} \\
\enddata
\tablerefs{(1) \cite{balmer2025}, (2) \cite{xuan2026b} (3) \cite{matthews2026}, (4) \cite{sanghi2026}, (5) This work }
\tablenotetext{1}{Spectral fit C/O ratios used}
\tablenotetext{2}{[S/H] for AF Lep b used as proxy for its atmospheric metallicity}
\end{deluxetable*}

\section{Conclusions \label{sec:conclusions}}
In this work, we perform detailed characterization for the cold imaged planet host stars AF Lep A and $\epsilon$ Indi A, measuring the stellar atmospheric properties and subsequently using those for the measurement of elemental abundances. We utilize two different approaches to determine the $T_{\rm eff}$ and $\log{g}$: 1) a spectroscopic approach along the lines of previous work \citep{aneesh26}, and 2) an evolutionary model approach utilizing estimates of stellar age, stellar metallicity, and measured Tycho-2+2MASS photometry. We find differences of up to $\sim260\,$K in the $T_{\rm eff}$ measurements, enabling us to account for the effects of temperature systematics while determining elemental abundances for the host stars.

We measure abundances of C, O, and S of AF Lep A and $\epsilon$ Indi using both spectral fitting and equivalent width approaches, and the abundances of 13 other elements (Na, Mg, Si, K, Ca, Sc, Ti, Cr, Mn, Fe, Zn, Y) using only the equivalent width method. Both approaches give solar C/O, C/S, and O/S abundance ratios (within 1.5$\sigma$) for both AF Lep A and $\epsilon$ Indi A. Among all species, those deviating from solar at $>2\sigma$ include Sc \textsc{ii} and Y \textsc{ii} for AF Lep A, and C \textsc{i}, K \textsc{i}, Sc \textsc{ii}, Ti \textsc{i}, Ti \textsc{ii}, Cr \textsc{ii}, Ni \textsc{i}, Zn \textsc{i}, and Y \textsc{ii} for $\epsilon$ Indi A. Additionally, we compare our abundance values for the latter with those measured by \cite{2022AJ....164...87K}, finding agreement within 1.5$\sigma$ for all elements. 

Lastly, we compare the derived stellar abundances to those of the planets around them. The super-stellar metallicities for both AF Lep b and $\epsilon$ Indi Ab relative to their host stars strongly support planet-like formation via the core-accretion pathway for these planets.

\section*{acknowledgements}
\noindent The authors would like to thank Henrique Reggiani for his support in obtaining these observations. 
\noindent J.W.X is grateful for support from the Heising-Simons Foundation 51 Pegasi b Fellowship (grant \#2025-5887).
\noindent This research was supported in part through the computational resources and staff contributions provided for the Quest high performance computing facility at Northwestern University which is jointly supported by the Office of the Provost, the Office for Research, and Northwestern University Information Technology.
\noindent This research has made use of the NASA Exoplanet Archive, which is operated by the California Institute of Technology, under contract with the National Aeronautics and Space Administration under the Exoplanet Exploration Program. Any opinions, findings, conclusions, and/or recommendations expressed in this paper are those of the author(s) and do not reflect the views of the National Aeronautics and Space Administration. This work is based on observations obtained at the international Gemini Observatory, a program of NSF NOIRLab, which is managed by the Association of Universities for Research in Astronomy (AURA) under a cooperative agreement with the U.S. National Science Foundation on behalf of the Gemini Observatory partnership: the U.S. National Science Foundation (United States), National Research Council (Canada), Agencia Nacional de Investigaci\'{o}n y Desarrollo (Chile), Ministerio de Ciencia, Tecnolog\'{i}a e Innovaci\'{o}n (Argentina), Minist\'{e}rio da Ci\^{e}ncia, Tecnologia, Inova\c{c}\~{o}es e Comunica\c{c}\~{o}es (Brazil), and Korea Astronomy and Space Science Institute (Republic of Korea). The data were obtained under Program ID GS-2025B-Q-210. This work has made use of the VALD database, operated at Uppsala University, the Institute of Astronomy RAS in Moscow, and the University of Vienna. This work has made use of data from the European Space Agency (ESA) mission {\it Gaia} (\url{https://www.cosmos.esa.int/gaia}), processed by the {\it Gaia} Data Processing and Analysis Consortium (DPAC,
\url{https://www.cosmos.esa.int/web/gaia/dpac/consortium}). Funding for the DPAC has been provided by national institutions, in particular the institutions participating in the {\it Gaia} Multilateral Agreement. \\
This work was conducted at Northwestern University, which sits on the traditional homelands of the people of the Council of Three Fires, the Ojibwe, Potawatomi, and Odawa as well as the Menominee, Miami and Ho-Chunk nations. It was also a site of trade, travel, gathering and healing for more than a dozen other Native tribes and is still home to over 100,000 tribal members in the state of Illinois.

\facilities{Gemini:South}
\software{emcee \citep{emcee}, SMART \citep{2021ApJS..257...45H}, SME \& PySME \citep{valenti1996,piskunov2017}, MOOG \citep{1973ApJ...184..839S}, SciPy \citep{2020SciPy-NMeth}, NumPy \citep{harris2020array}, matplotlib \citep{Hunter:2007}, astropy \citep{astropy:2013, astropy:2018, astropy:2022}, corner \citep{corner}, pysynphot \citep{2013ascl.soft03023S}}

\appendix
Fits of the \textit{MARCS}-\textit{C/O} grid to the spectrum of $\epsilon$ Indi A and the corresponding posterior plots.
\begin{figure}
    \centering
    \includegraphics[width=0.9\textwidth]{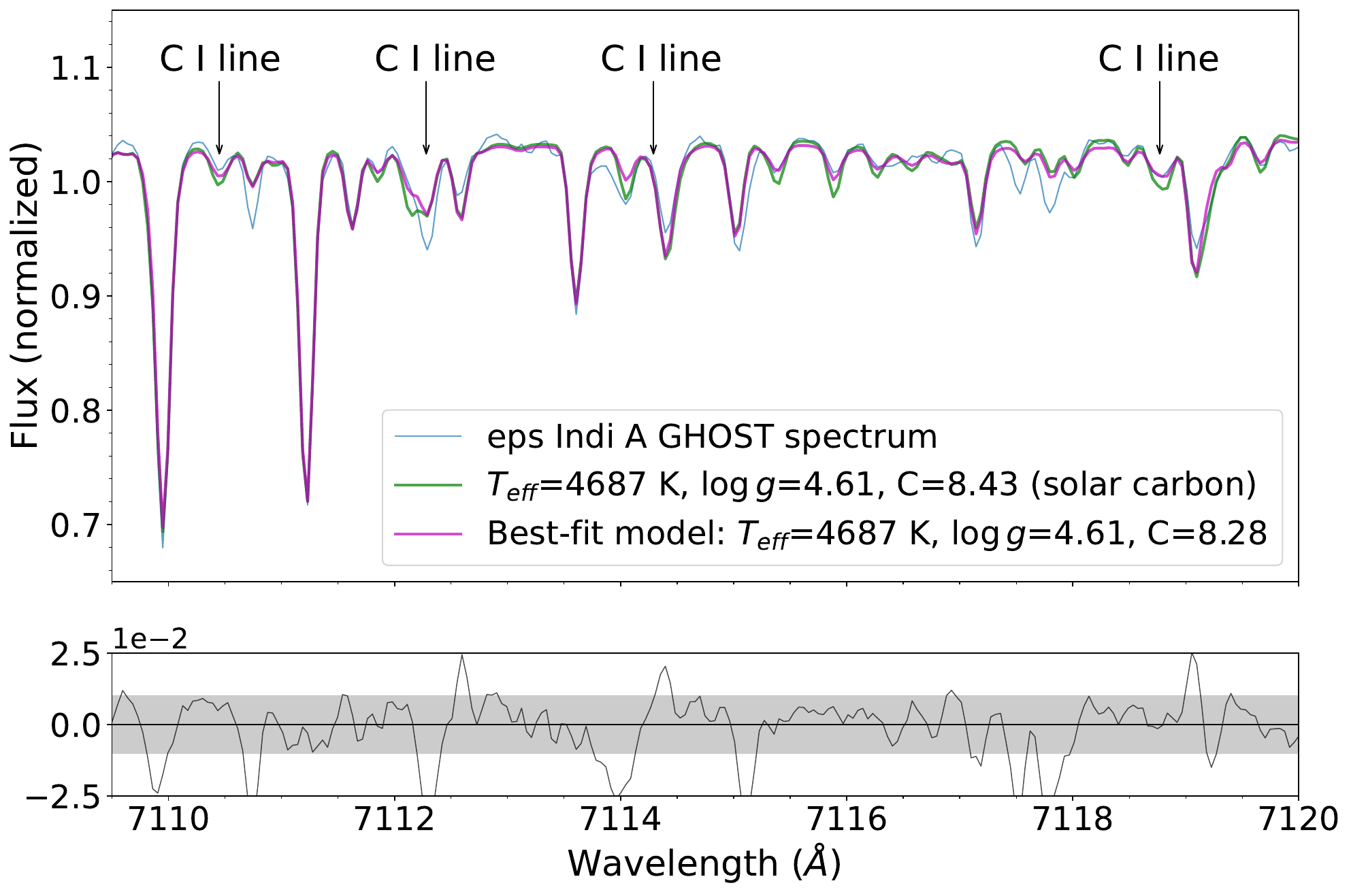}
    \caption{Best-fit \textit{MARCS-C/O} model (magenta) to the GHOST spectrum of $\epsilon$ Indi A (cyan), shown for C \textsc{i} lines between 7110--7119 \r{A}. The best-fit model has carbon abundance $\epsilon_C$ = 8.28. The residuals between the data and the model are plotted in black and other noise limits are shown in gray. Also shown is a model with the same $T_{\rm eff}$ and $\log{g}$ but with solar carbon abundance $\epsilon_C$ = 8.43 (green). The solar abundance model is a worse fit to the C \textsc{i} lines (especially the 7119 \r{A} line) compared to our best-fit model.}
    \label{fig:epsindspec}
\end{figure}

\begin{figure}
    \centering
    \includegraphics[width=0.8\textwidth]{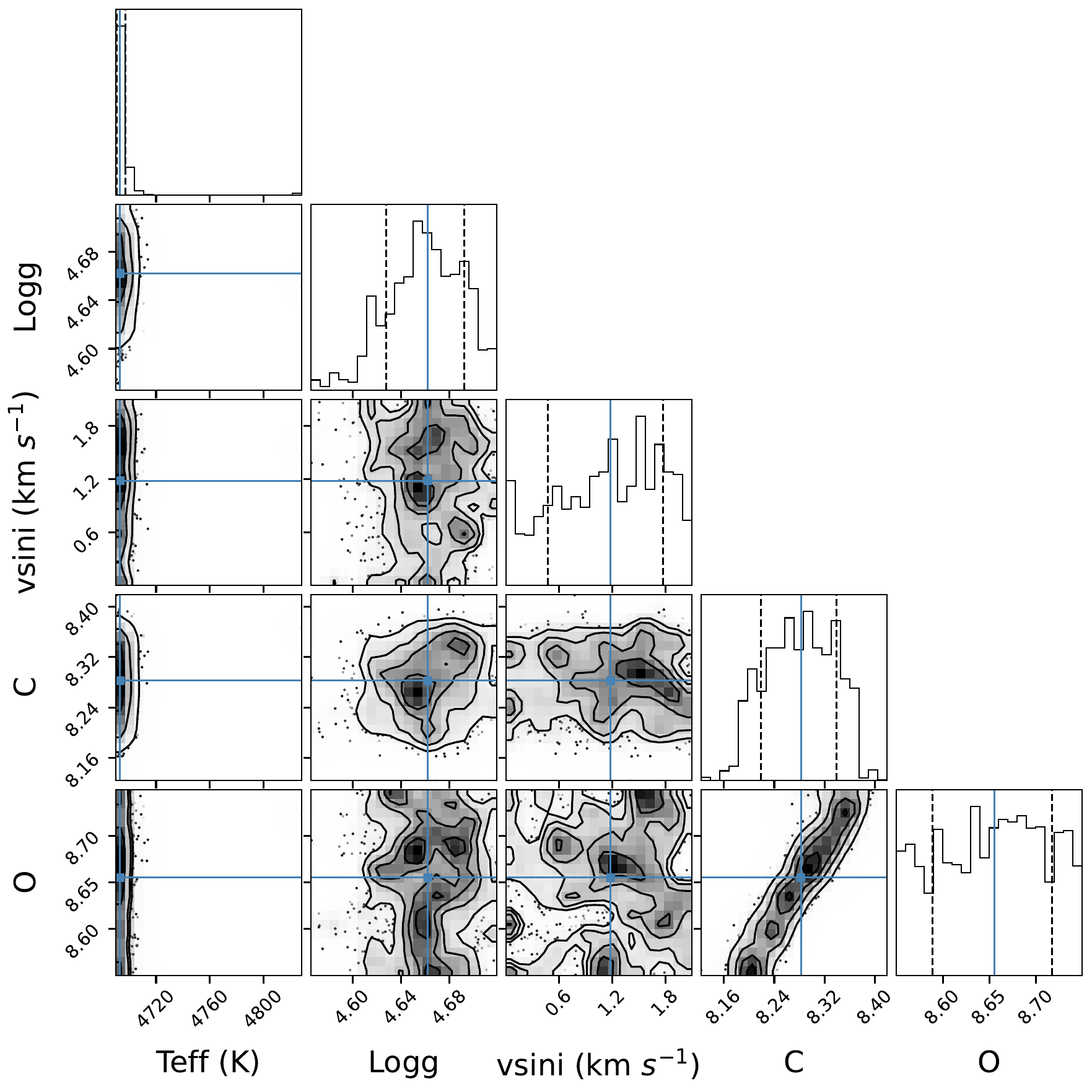}
    \caption{Posteriors for \textit{MARCS}--\textit{C/O} grid fit to spectral order with C \textsc{i} lines between 7110--7119 \r{A}. The oxygen abundance is fixed to $\epsilon_O$ = 8.65 $\pm$ 0.10. The marginalized posteriors are shown along the diagonal. The blue lines represent the 50 percentile, and the dotted lines represent the 16 and 84 percentiles. The subsequent covariances between all the parameters are in the corresponding 2-D histograms. We obtain a best-fit $\log{\epsilon_C}$ = 8.28 $\pm$ 0.06.}
    \label{fig:epsindcorner}
\end{figure}

\bibliography{main}
\end{document}